\documentclass[11pt,twoside]{article}
\usepackage{asp2010}

\resetcounters

\begin{document}

\title{LEDDB: LOFAR Epoch of Reionization Diagnostic Database}
\author{O. Martinez-Rubi$^1$, V. K. Veligatla$^1$, A. G. de Bruyn$^{1,2}$, P. Lampropoulos$^2$, A. R. Offringa$^1$,  V. Jelic$^{1,2}$, S. Yatawatta$^2$, L. V. E. Koopmans$^1$, and S. Zaroubi$^1$
\affil{$1$ Kapteyn Astronomical Institute, University of Groningen, \\
PO Box 800, 9700 AV Groningen, the Netherlands\\
$2$ ASTRON, the Netherlands Institute for Radio Astronomy, \\
PO Box 2, 7990 AA Dwingeloo, the Netherlands}
}

\begin{abstract}
One of the key science projects of the Low-Frequency Array (LOFAR) is the detection of the cosmological signal coming from the Epoch of Reionization (EoR). Here we present the LOFAR EoR Diagnostic Database (LEDDB) that is used in the storage, management, processing and analysis of the LOFAR EoR observations. It stores referencing information of the observations and diagnostic parameters extracted from their calibration. This stored data is used to ease the pipeline processing, monitor the performance of the telescope and visualize the diagnostic parameters which facilitates the analysis of the several contamination effects on the signals. It is implemented with PostgreSQL and accessed through the psycopg2 python module. We have developed a very flexible query engine, which is used by a web user interface to access the database, and a very extensive set of tools for the visualization of the diagnostic parameters through all their multiple dimensions.
\end{abstract}

\section{Introduction}

The Low-Frequency Array (LOFAR) is an antenna array that observes at low radio frequencies (10 - 240 MHz). It consists of about 70 stations spread around Europe that combine their signals to form an interferometric aperture synthesis array \citetext{van Haarlem et al. in preparation}. The LOFAR Epoch of Reionization (EoR) experiment is one of the key science projects (KSP) of LOFAR. It aims to study the redshifted 21-cm line of neutral hydrogen from the Epoch of Reionization \citetext{de Bruyn et al. in preparation}. There are many challenges that need to be overcome in order to meet this goal including strong astrophysical foreground contamination, ionospheric distortions, complex instrumental response and different types of noise. The very faint signals from neutral hydrogen require hundreds of hours of observation thereby accumulating petabytes of data. To diagnose and monitor the various instrumental and ionospheric parameters, as well as manage the data, we have developed the LEDDB (LOFAR EoR Diagnostic Database). Its main tasks and uses are:

\begin{itemize}
\item To store referencing information of the observations, mainly the locations of the data but also other indexing information.
\item To store diagnostic parameters of the observations extracted through calibration.
\item To facilitate efficient data management and pipeline processing.
\item To monitor the performance of the telescope as a function of date.
\item To visualize the diagnostic parameters. For example we can observe the complex gain of all the stations as a function of time and frequency to visualize ionospheric distortion affecting large part of the array.
\end{itemize}

\section{Data flow}
The data from the stations is sent to the Central Processing Facility (CEP) located in Groningen (the Netherlands), where it is correlated among other processing steps. Afterwards, the data is stored in the Long Term Archive (LTA) in Groningen. From the LTA we copy the data to the LOFAR EoR CPU/GPU cluster, also in Groningen, where we process it with the LOFAR EoR pipeline. The LEDDB takes care of storing the locations of the data both in the LTA and the LOFAR EoR cluster. It also stores all the diagnostic data produced by the pipeline. Since we can not keep all the data in the LOFAR EoR cluster, we must archive it in the LTA but thanks to the LEDDB we retain access to all its diagnostic information.

\section{Database definition} 

\begin{figure}[!ht]
  \centering
    \includegraphics[scale=0.077]{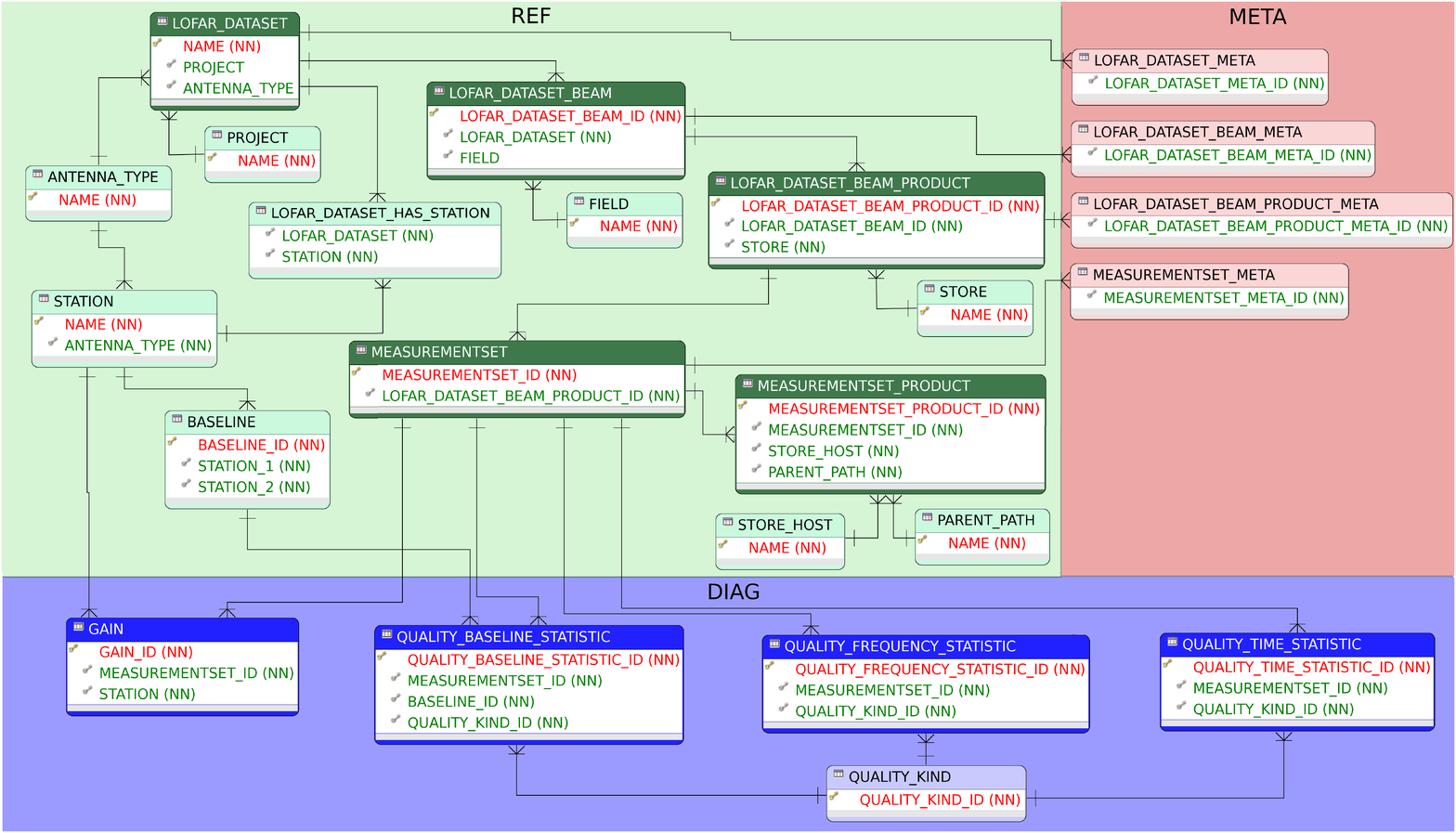} 
  \caption{Entity-Relationship diagram of the LEDDB. Only table names and key columns are shown.}
  \label{fig:leddber}
\end{figure}

The LEDDB is implemented with PostgreSQL and accessed through a python interface provided by the psycopg2 module. It is part of a research project with still evolving requirements, so one of the key points of the design was to make it flexible enough to meet new requirements such as the addition of new diagnostic parameters. The content of the database is categorized under three different blocks: the referencing information, the diagnostic data and the meta-data. In figure \ref{fig:leddber} we show the Entity-Relationship diagram of the database with its blocks, the tables involved and their relationships.

(1) The referencing information block (\textit{``REF''} in figure \ref{fig:leddber}) contains five primary tables: \textit{LOFAR\-\_DATA\-SET} (LDS), \textit{LOFAR\-\_DATA\-SET\-\_BEAM} (LDSB), \textit{LOFAR\-\_DATA\-SET\-\_BEAM\-\_PRODUCT} (LDSBP), \textit{MEA\-SU\-RE\-MENT\-SET} (MS) and {MEA\-SU\-RE\-MENT\-SET\-\_PRODUCT} (MSP). They contain information about the observations: their names, date and time information, the pointed fields and other indexing information. They also store the locations of the data, i.e., the host and cluster the data is in and the path to the files. The rest of tables in this block are the secondary tables which are only used to ease the selection on the primary ones.

(2) The diagnostic data block (\textit{``DIAG''} in figure \ref{fig:leddber}) contains the diagnostic parameters related to the observations. There are four primary tables: the \textit{GAIN} table and three \textit{QUALITY} tables. They store the gain solutions of the stations and baseline-based, frequency-based and time-based statistic parameters of the data. There is also a secondary table in this block called QUA\-LI\-TY\-\_KIND.

(3) Finally the meta-data block (\textit{``META''} in figure \ref{fig:leddber}) stores information regarding the relationships of the referencing section and the diagnostic data. Each one of the referencing tables is joined with each related meta-data table.

The LEDDB can generate a \textit{RefFile} or a \textit{DiagFile}. A \textit{RefFile} is a file containing locations of data related to the observations. This file is used in the LOFAR EoR pipeline processing tasks. On the other hand, a \textit{DiagFile} contains references to diagnostic data in the LEDDB.

\section{Diagnostic data analysis}
\begin{figure}[!ht]
  \centering
    \includegraphics[scale=0.19]{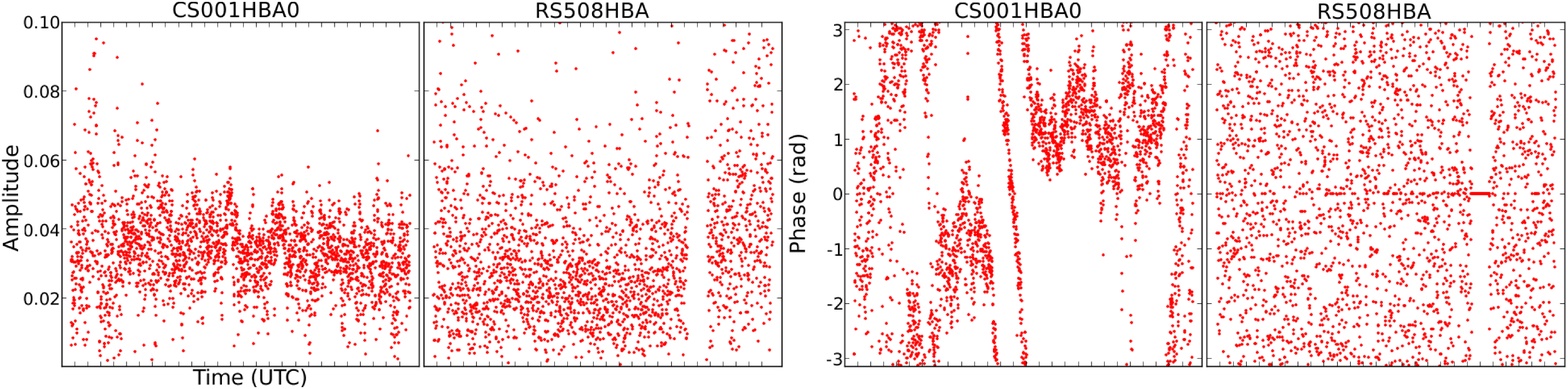} 
  \caption{Gain as a function of time of one of the polarization auto-correlations of two different stations at 138 MHz for the observation L60639 (Elais field). Note the phase difference between a core station (CS001HBA0) and a remote station (RS508HBA), mainly caused by the ionosphere.}
  \label{fig:gain}
\end{figure}

The diagnostic data can have multiple dimensions: Time, frequency, baseline (interferometer), station, polarization correlations and other ones depending on the situation. In general they are complex numbers. We provide plotting and animation tools implemented with matplotlib to analyse such multi-dimensional data. In figure \ref{fig:gain} we show an example of one of the produced plots.

\section{Query engine and User Interface}

\begin{figure}[!ht]
  \centering
    \includegraphics[scale=0.24]{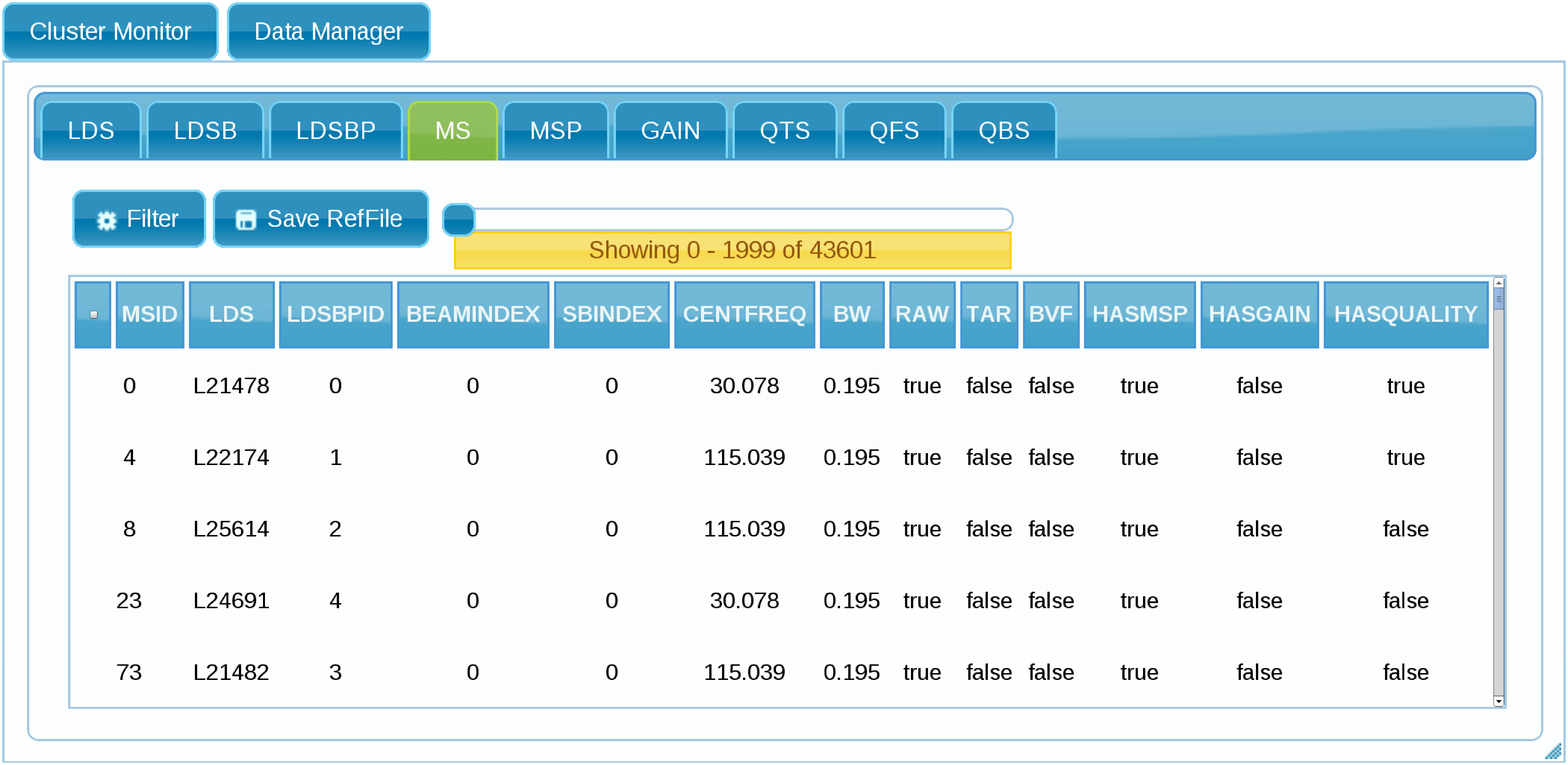} 
  \caption{Snapshot of the web UI. Each tab in the UI represents a primary table in the database.}
  \label{fig:webui}
\end{figure}

The query engine is a python API which provides fast and flexible access to the database.
We use a python based web server (cherrypy) to interface with the query engine. The client-side user interface (UI) in the web page is implemented with JQueryUI framework. In figure \ref{fig:webui} we show a snapshot of the web UI.

We estimate that 10 terabytes of diagnostic data will be stored in the LEDDB for the full LOFAR EoR KSP (currently it is 75 gigabytes). In addition to the size challenge, the number of rows of some of the tables is the most important aspect to be taken into account in the design of the database and its query engine, and it is actually the main bottleneck in the queries. We have managed to provide a fast access thanks to efficient table indexing, the minimization of the number of join operations and the use of persistent connections eased by the session handling provided by the cherrypy framework. 

The query engine provides functionality to sort, filter by column values and by selection in primary and secondary tables. This is used by the UI to provide a very extensive set of options for accessing the data.

The UI allows the user to create both \textit{RefFiles} and \textit{DiagFiles}. Besides, this UI can be used to launch pipeline jobs with a \textit{RefFile} and directly plot diagnostic data with a \textit{DiagFile}.

\section{Future developments}
We will focus on minimizing the access times while the database is growing and improving the tools to analyse the diagnostic parameters. Possibly new diagnostic parameters will added. There is also a plan to migrate the database to a new server specially designed for its purpose. 

\acknowledgements We are indebted to Eite Tiesinga for his assistance in all the matters related to the LOFAR EoR cluster. We also thank the rest of the LOFAR EoR group core members for their contribution. \newline (\textit{www.astro.rug.nl/eor/people/core-members}).

\bibliography{author}

\end{document}